\documentclass{svproc}
\usepackage{url}
\usepackage{amssymb}
\usepackage{amsmath}
\usepackage{epsfig}
\usepackage{slashed}
\usepackage{xcolor}

\usepackage{cite}
\usepackage[sectionbib,numbers,super,square,sort&compress]{natbib}
\usepackage[linktocpage]{hyperref}
\hypersetup{colorlinks=true,citecolor=red,linkcolor=red,urlcolor=MidnightBlue}
\usepackage[caption=false]{subfig}

\begin{document}
\mainmatter              
\title{LHC bounds on R$_{D^{(*)}}$ motivated Leptoquark models}
%
%
\author{Arvind Bhaskar\inst{1}, Tanumoy Mandal\inst{2},
Subhadip Mitra\inst{1}, Cyrin Neeraj\inst{1}\thanks{Speaker}, Swapnil Raz\inst{1} }
\authorrunning{A. Bhaskar et al.} 
%
%
\institute{Center for Computational Natural Sciences and Bioinformatics,
International Institute of Information Technology, Hyderabad 500 032, India.\\
\and
Indian Institute of Science Education and Research Thiruvananthapuram, Vithura, Kerala, 695 551, India.\\
\email{arvind.bhaskar@research.iiit.ac.in, cyrin.neeraj@research.iiit.ac.in, tanumoy@iisertvm.ac.in, subhadip.mitra@iiit.ac.in, swapnil.raz@research.iiit.ac.in}}

\maketitle              

\begin{abstract}
Most of the popular explanations of the observed anomalies in the semileptonic $B$-meson decays involve TeV scale Leptoquarks (LQs). Among the various possible LQ models, two particular LQs -- $S_{1}(3, 1, 1/3)$ and $U_{1}(3, 1, 2/3)$ seem to be most promising. Here, we use current LHC data to constrain the $S_{1}(3, 1, 1/3)$ and $U_{1}(3, 1, 2/3)$ parameter spaces relevant for the $R_{D^{(*)}}$ observables. We recast the latest ATLAS $\tau\tau$ resonance search data to obtain new exclusion limits. For this purpose, we consider both resonant (pair and single productions) and non-resonant ($t$-channel LQ exchange) productions of these LQs at the LHC. For the limits, the most dominant contribution comes from the (destructive) interference of the non-resonant production with Standard Model backgrounds. The combined contribution from the pair and inclusive single production processes~\cite{Mandal:2015vfa} is less prominent but non-negligible.
The limits we get are independent  and competitive to other known bounds. For both the models, we set limits on $R_{D^{(*)}}$ motivated couplings~\cite{Mandal:2018kau}~\cite{Bhaskar:2021pml}.
\end{abstract}
\section{Introduction}
Leptoquarks (LQs) are colored bosons (scalar or vector) that can couple with Standard Model (SM) leptons and quarks. Some of them are well-suited candidates to account for the observed anomalies in the decays of the $B$-meson. At present, the data for the $R_{D^{(*)}}$ observables, defined as,
\begin{align}
R_{D^{(*)}}  =  \dfrac{{\mathcal {B}}(B\rightarrow D^{(*)}\tau\bar\nu)}{{\mathcal {B}}(B\rightarrow D^{(*)}\hat{\ell}\bar\nu)} \label{eq:anomalies}
\end{align}
show a combined excess of $\sim 3.1\sigma$~\cite{Amhis:2016xyh} than the Standard Model predictions. Here we consider two weak-singlet LQs -- the scalar $S_1({\bf 3}, {\bf 1}, -1/3)$ and the vector $U_{1}({\bf 3}, {\bf 1}, 2/3)$. Both are popular candidates for explaining the $R_{D^{(*)}}$ anomalies in the literature. 

Our objective in this study is to obtain precise exclusion limits on the parameter spaces of these LQs from LHC data that are complimentary to and independent of other known flavor bounds. We do this by recasting the latest LHC dilepton search results in the $\tau\tau$ channel~\cite{Aaboud:2017sjh}. We show that LHC is sensitive to the model parameters like the cross generational couplings that are needed to account for $R_{D^{(*)}}$. 

\section{The $S_1$ and $U_1$ Models, Relevant Processes}
 The Lagrangian for $S_1$ model looks like,~\cite{Mandal:2018kau}
\begin{equation}
    \label{eq:Lcompact_S1}
	\mathcal{L}\supset \left[ \lambda_{33}^L\,\bar{Q}^c_3\left(i\tau_2\right) L_3 + \lambda_{23}^L\,\bar{Q}^c_2\left(i\tau_2\right)L_3 + \lambda_{23}^R\,\bar{c}^c\tau_R \right]{\mathcal{S}^\dagger_1} + H.c. 
\end{equation}
where $Q_\alpha(L_\alpha)$ denotes the $\alpha$-th generation quark (lepton) doublet, $\lambda^X_{ab}$ denotes the coupling of $\mathcal{S}_1$ with an $a^{\rm th}$ generation quark and a lepton of generation $b$ with chirality $X$. For the $R_{D^{(*)}}$ observables, the $S_1$ is required to couple with $b\nu$ and $c\tau$. This makes the scenario with only $\lambda^{R}_{23}$ coupling non-zero inconsequential. Here, for illustration, we choose two minimal scenarios with either $\lambda^{L}_{23}\ne0$ or $\lambda^{L}_{33}\ne0$. In these scenarios, the one of the desired couplings are generated from the other via CKM mixing of the quarks.
 For $U_1$, the interaction terms are given as,~\cite{Bhaskar:2021pml}
\begin{equation}
    \label{eq:GenLagU1}
    \mathcal{L} \supset
    \left[\lambda_{23}^{L}~\bar{Q}_{2}\gamma_{\mu}P_{L}L_{3} + \lambda_{33}^{L}~\bar{Q}_{3}\gamma_{\mu}P_{L}L_{3} + \lambda_{33}^{R}~\bar{b}~\gamma_{\mu}\tau_{R}\right]U^{\mu}_1 + H.c. 
    \end{equation}
Here the essential couplings for the $R_{D^{(*)}}$ observables are $c\nu U_1$ and $b\tau U_1$. We choose similar minimal scenarios (as $S_1$) to inspect the LHC bounds.

As mentioned above, we recast the latest $\tau\tau$ search data from ATLAS~\cite{Aaboud:2017sjh}. Here, we illustrate how various processes -- both resonant (single production, pair production) and non-resonant ($t-$channel lepton exchange) -  contribute to the $\tau\tau$ final state for $S_1$ (which puts no additional restriction on the extra jets). The case for $U_1$ can be argued similarly.
When $\lambda_{23}^L$ is non-zero, pair production of $S_1$ contributing to $\tau\tau$ final state is 
\begin{equation}
    pp\to
    \mathcal{S}_1\mathcal{S}_1 \to c\tau\:c\tau \equiv \tau\tau+2j 
\end{equation}
Single productions would also lead to the same final state~\cite{Mandal:2015vfa}. The processes that contribute to the $\tau\tau$ final states are 
\begin{equation}
    pp\to\left\{\begin{array}{lclcl}
    \mathcal{S}_1\ \tau &\to & \tau j\:\:\ \tau \\
    \mathcal{S}_1\ \tau j &\to & \tau j\:\:\ \tau j \label{eq:match_tautau}\\
    \mathcal{S}_1\ \tau jj &\to & \tau j\:\:\ \tau jj
    \end{array}\right\} 
\end{equation}
These processes are combined using MLM matching to prevent double-counting. 
The $\tau\tau$ final state can also be produced via non-resonant processes ($t-$channel exchange of $S_1$). They interfere destructively with the SM background processes and have significantly higher cross sections than the resonant processes considered, especially for heavier LQs.

Similarly, we find processes contributing to $\tau\tau$ final states for each coupling and combine them systematically to obtain the total contributions from the model. We then recast it with the LHC $\tau\tau$ search data to obtain the latest bounds on each of the model parameters. For details of the recast, see Refs~\cite{Mandal:2018kau,Bhaskar:2021pml}.

\begin{figure}[!t]
\centering\captionsetup[subfigure]{labelformat=empty}
\subfloat[\quad(a)]{\includegraphics[width=0.45\textwidth]{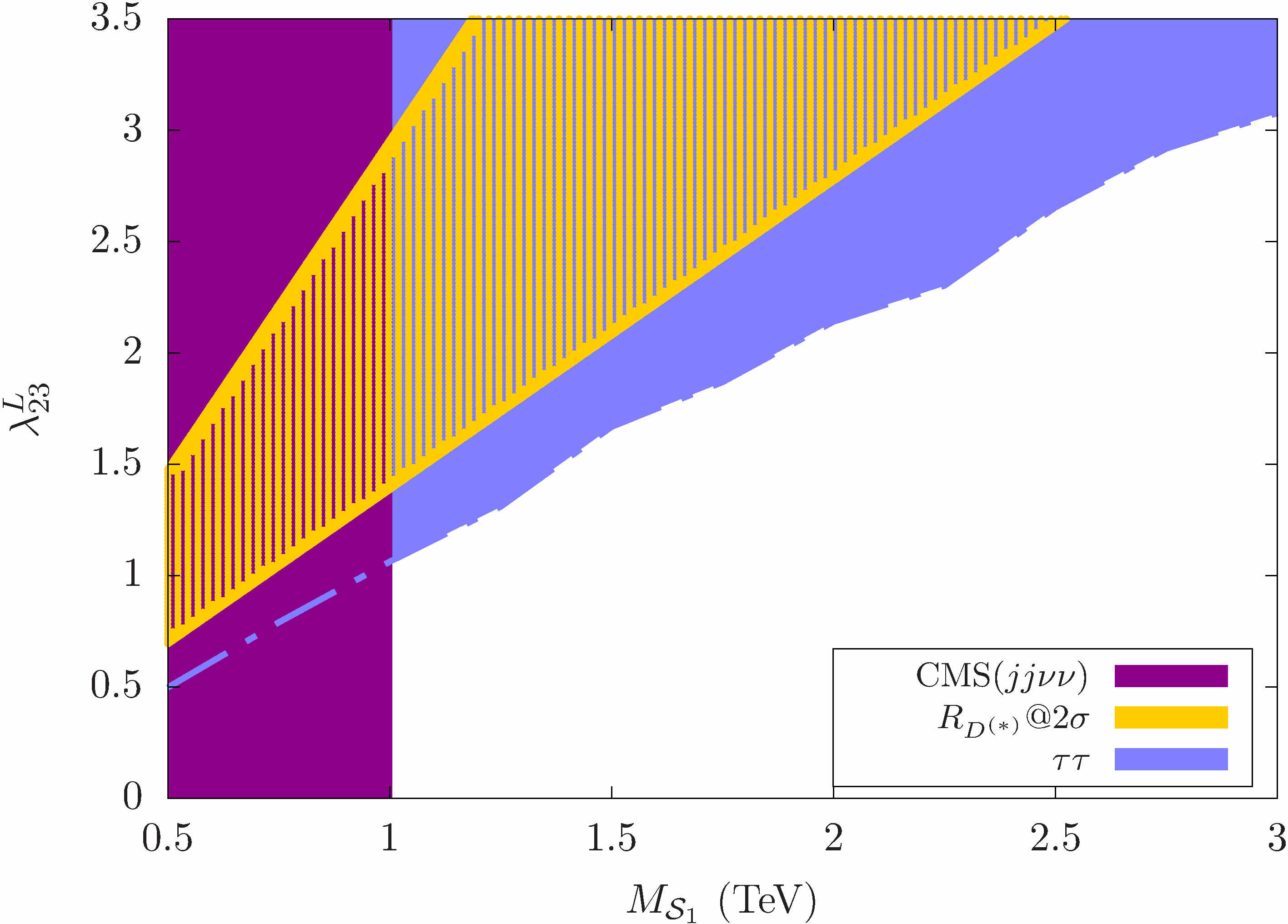}\label{fig:Slm23l}}\quad\quad
\subfloat[\quad(b)]{\includegraphics[width=0.45\textwidth]{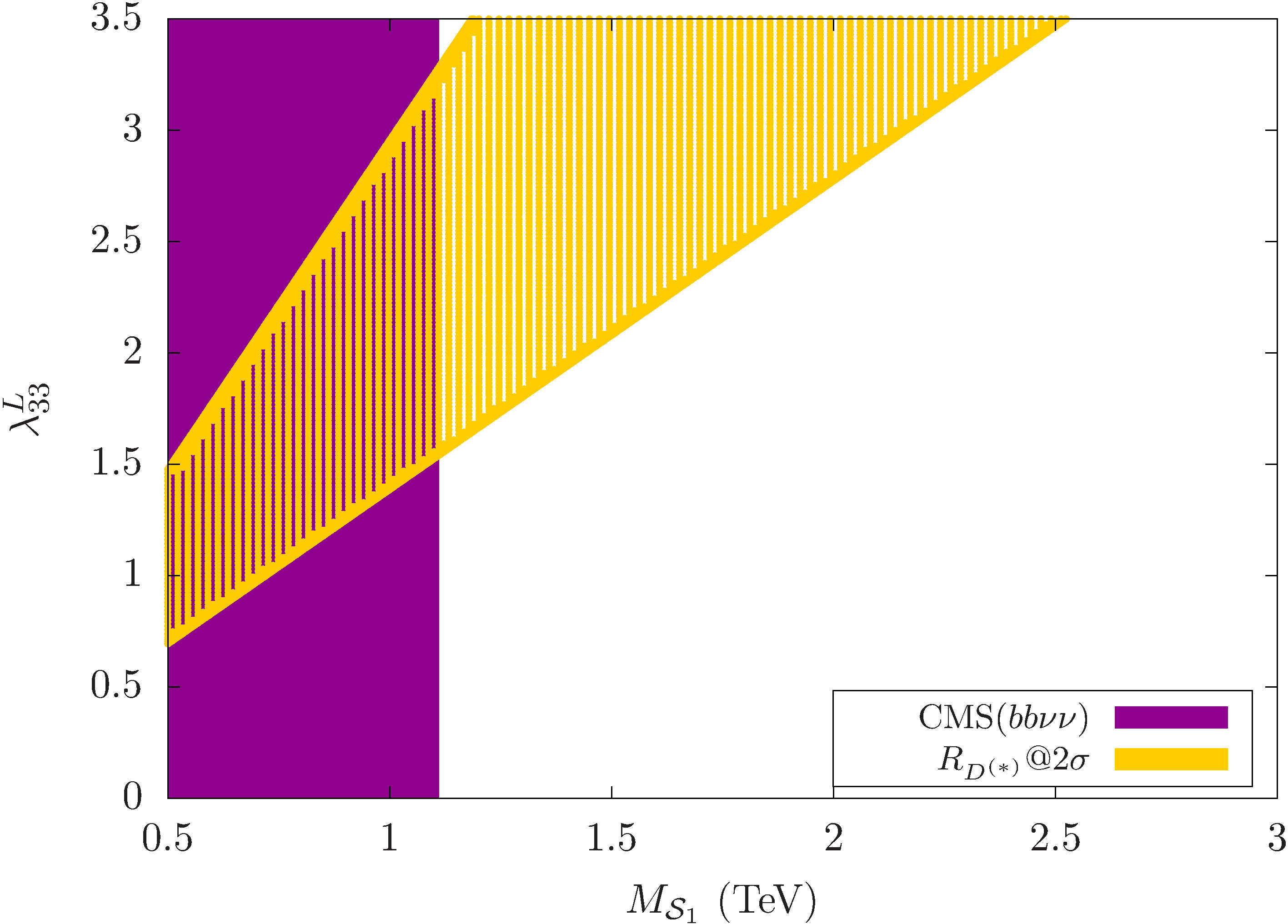}\label{fig:Slm33l}}\\
\subfloat[\quad(c)]{\includegraphics[width=0.45\textwidth]{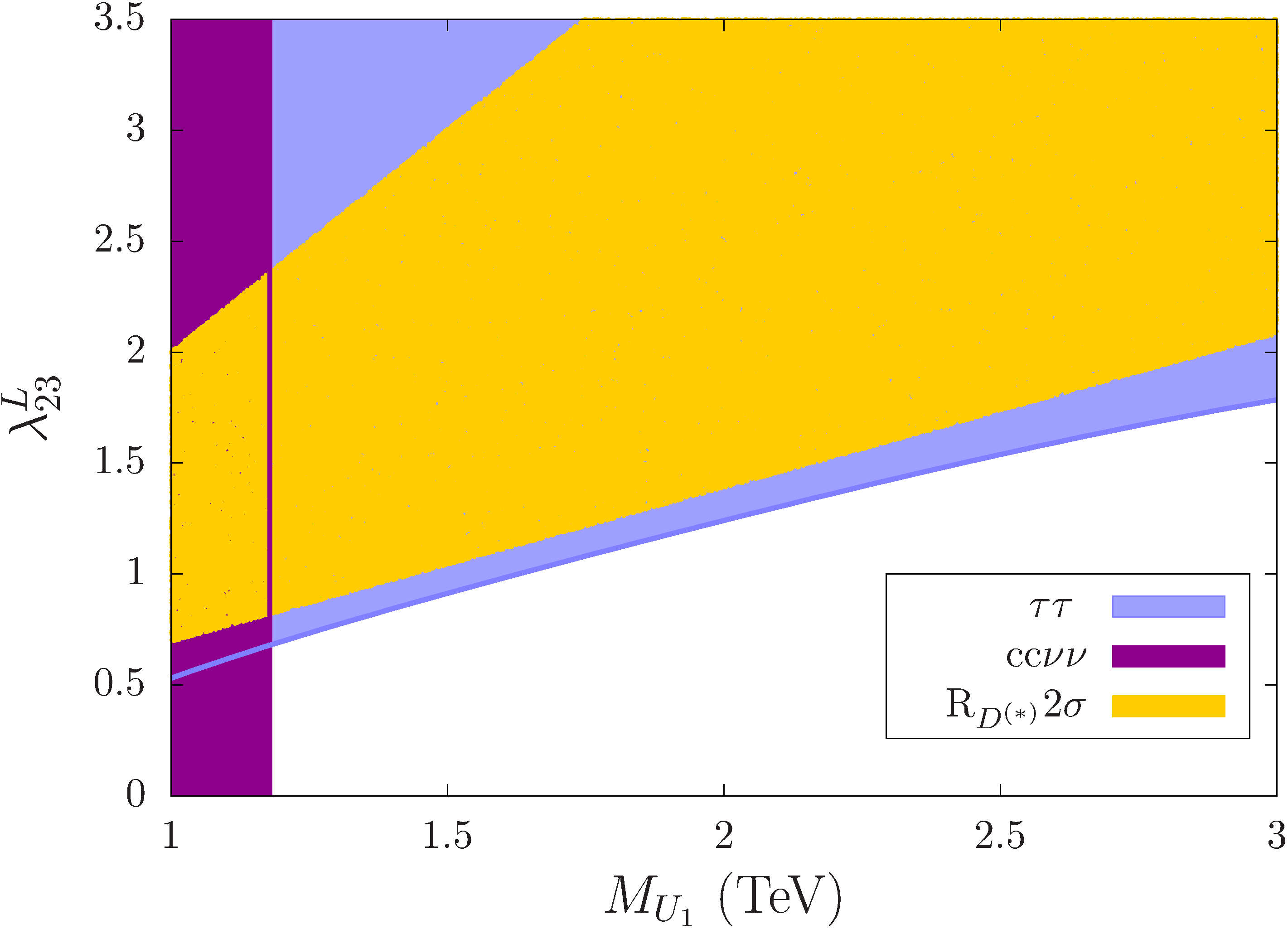}\label{fig:vlm23l}}\quad\quad
\subfloat[\quad(d)]{\includegraphics[width=0.45\textwidth]{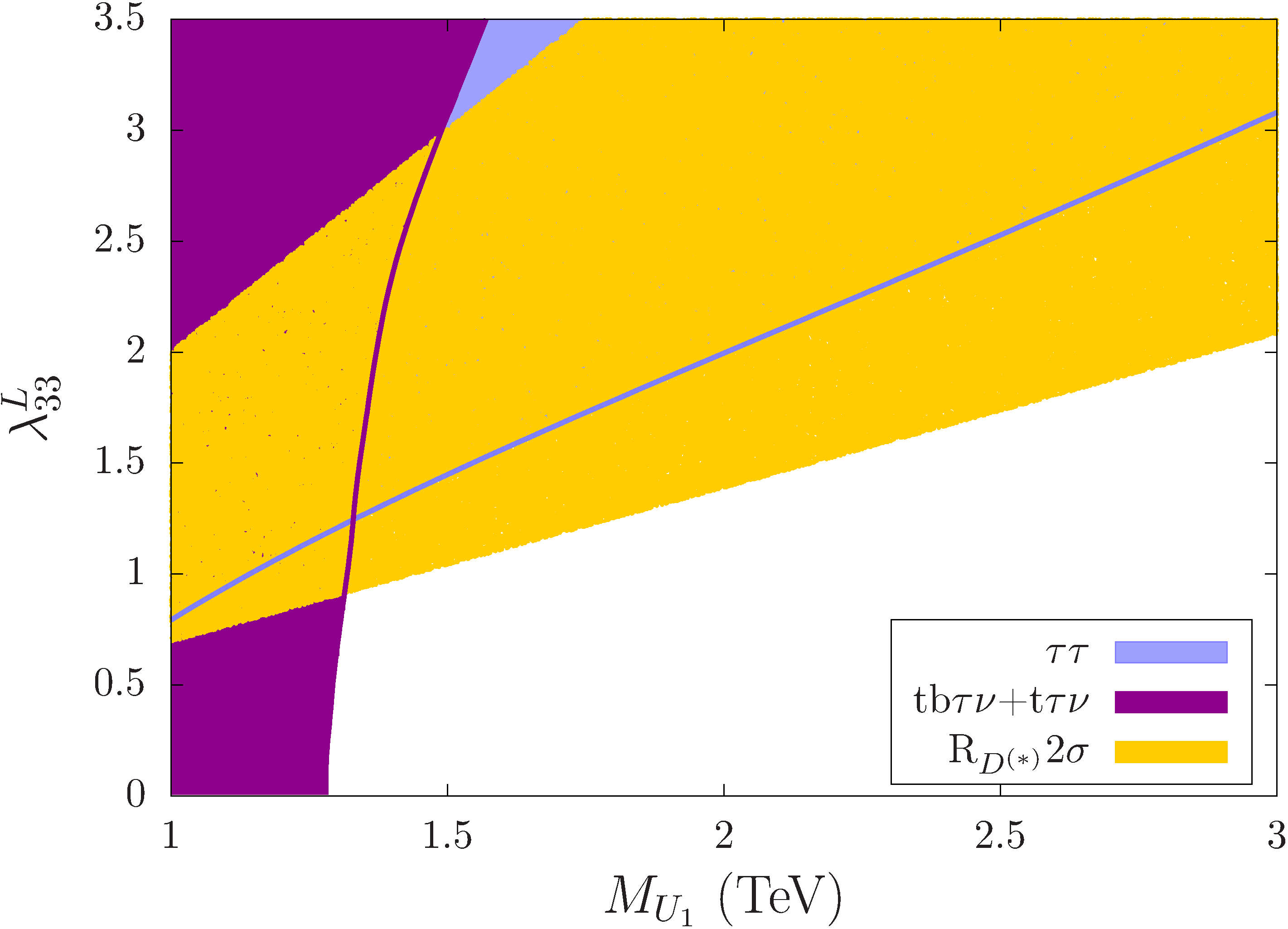}\label{fig:vlm33l}}
\caption{The $2\sigma$ exclusion limits from the LHC and the preferred regions by the $R_{D^{(*)}}$ in the M$_{LQ}$-$\lambda$ parameter space. The violet regions show the $2\sigma$ bounds from the LHC $\tau\tau$ data. The yellow region depicts the $R_{D^{(*)}}$ favoured regions. The magenta regions are obtained by recasting the CMS pair production search data~\cite{Sirunyan:2018kzh}. Plots on the top row show the bounds on $S_1$ parameter space - (a) Only $\lambda_{23}^L\ne0$, (b) Only $\lambda_{33}^L\ne0$. Bottom row plots show the bounds on $U_1$ parameter space - (c) Only $\lambda_{23}^L\ne0$, (d) Only $\lambda_{33}^L\ne0$. The magenta region here is obtained by recasting the CMS pair and single production process \cite{Sirunyan:2020zbk}.} 
\label{fig:LM_Ms}
\end{figure}
\section{Results}

In Fig.~\ref{fig:LM_Ms}, we show the exclusion bounds obtained from the LHC dilepton data along with the R$_{D^{(*)}}$ favorable regions. We have also shown the  limits from direct searches (obtained after recast, where necessary). The purple region shows the parameter space excluded with a CL of $95$\% by the $\tau\tau$ data. In Fig.~\ref{fig:Slm33l}, since the LHC is insensitive to the $\lambda^{L}_{33}$ coupling, the bounds on this scenario come only from the pair production searches. In Figs.~\ref{fig:Slm23l} and~\ref{fig:vlm23l}, the $R_{D^{(*)}}$ favored regions is completely ruled out by the LHC data. In the case of Fig.~\ref{fig:vlm33l}, there is a minor region which is still allowed by the LHC data and also explains the R$_{D^{(*)}}$ observables.

\section{Conclusions, Remarks}
\label{conclusion}
We see that the LHC is indeed sensitive to model specific parameters and hence, we use that fact to put constraints on the parameter space of the model specific couplings. The results shown above are the most updated and precise limits on the $M_{LQ}$-$\lambda$ parameter space of both $S_1$, $U_1$ models for the parameters considered. We find that the non-resonant interferes destructively with the SM and therefore has the most dominant contribution compared to all processes considered here, due to the SM background being large. In the minimal scenarios $R_{D^{(*)}}$, we find that almost most of favorable regions are under stress by the limits from the LHC data. 

 The minimal scenarios considered above form only a subset of the the possible coupling choices. We have considered scenarios with multiple couplings in Refs~\cite{Mandal:2018kau,Bhaskar:2021pml}. There we demonstrated a generic method to combine contributions from multiple couplings systematically to obtain  parameter regions allowed by the dilepton data.\\

\label{Acknowledgments}
\noindent {\bf {Acknowledgments:}}
A.B. and S.M. acknowledge support from the Science and
Engineering Research Board, India, under grant number ECR/2017/000517. T.M. is supported by the intramural grant from IISER-TVM. C.N. is supported by the DST-Inspire Fellowship. 

%
%
\end{document}